\documentclass[conference,10pt,a4paper]{IEEEtran}
\usepackage{amsmath}
\usepackage{amssymb}
\usepackage[T1]{fontenc}
\usepackage{times}
\usepackage{graphicx}
\usepackage{multirow}
\usepackage[none]{hyphenat}
\usepackage{float}
\usepackage{subfig}
\usepackage{xurl}

\usepackage{algorithmic}
\usepackage{textcomp}
\usepackage{xcolor}
\usepackage{booktabs}
\usepackage{tabularx}
\usepackage{array}

\makeatletter

\def\@maketitle{\newpage
\bgroup\par\addvspace{0.5\baselineskip}\centering%
\ifCLASSOPTIONtechnote
   {\bfseries\large\@IEEEcompsoconly{\sffamily}\@title\par}\vskip 1.3em{\lineskip .5em\@IEEEcompsoconly{\sffamily}\@author
   \@IEEEspecialpapernotice\par{\@IEEEcompsoconly{\vskip 1.5em\relax
   \@IEEEtitleabstractindextextbox{\@IEEEtitleabstractindextext}\par
   \hfill\@IEEEcompsocdiamondline\hfill\hbox{}\par}}}\relax
\else
   \vskip0.2em{\EuMWtitlesize\ifCLASSOPTIONtransmag\bfseries\LARGE\fi\@IEEEcompsoconly{\sffamily}\@IEEEcompsocconfonly{\normalfont\normalsize\vskip 2\@IEEEnormalsizeunitybaselineskip
   \bfseries\Large}\@title\par}\vskip1.0em\par
   \ifCLASSOPTIONconference%
      {\@IEEEspecialpapernotice\mbox{}\vskip\@IEEEauthorblockconfadjspace%
       \mbox{}\hfill\begin{@IEEEauthorhalign}\@author\end{@IEEEauthorhalign}\hfill\mbox{}\par}\relax
   \else
      \ifCLASSOPTIONpeerreviewca
         {\@IEEEcompsoconly{\sffamily}\@IEEEspecialpapernotice\mbox{}\vskip\@IEEEauthorblockconfadjspace%
          \mbox{}\hfill\begin{@IEEEauthorhalign}\@author\end{@IEEEauthorhalign}\hfill\mbox{}\par
          {\@IEEEcompsoconly{\vskip 1.5em\relax
           \@IEEEtitleabstractindextextbox{\@IEEEtitleabstractindextext}\par\hfill
           \@IEEEcompsocdiamondline\hfill\hbox{}\par}}}\relax
      \else
         \ifCLASSOPTIONtransmag
           {\@IEEEspecialpapernotice\mbox{}\vskip\@IEEEauthorblockconfadjspace%
            \mbox{}\hfill\begin{@IEEEauthorhalign}\@author\end{@IEEEauthorhalign}\hfill\mbox{}\par
           {\vspace{0.5\baselineskip}\relax\@IEEEtitleabstractindextextbox{\@IEEEtitleabstractindextext}\vspace{-1\baselineskip}\par}}\relax
         \else
           {\lineskip.5em\@IEEEcompsoconly{\sffamily}\sublargesize\@author\@IEEEspecialpapernotice\par
           {\@IEEEcompsoconly{\vskip 1.5em\relax
            \@IEEEtitleabstractindextextbox{\@IEEEtitleabstractindextext}\par\hfill
            \@IEEEcompsocdiamondline\hfill\hbox{}\par}}}\relax
         \fi
      \fi
   \fi
\fi\par\addvspace{0.0\baselineskip}\egroup}

\def\EuMWtitlesize{\@setfontsize{\EuMWtitlesize}{24}{24pt}}
\def\EuMWauthorsize{\@setfontsize{\EuMWauthorsize}{11}{11pt}}
\def\EuMWaffilsize{\@setfontsize{\EuMWaffilsize}{10}{10pt}}
\def\EuMWcaptionsize{\@setfontsize{\EuMWcaptionsize}{9}{10pt}}
\def\EuMWbibsize{\@setfontsize{\EuMWbibsize}{8}{10pt}}

\def\@IEEEauthorblockNstyle{\EuMWauthorsize\@IEEEcompsocnotconfonly{\sffamily}\@IEEEcompsocconfonly{\large}}
\def\@IEEEauthorblockAstyle{\EuMWaffilsize\@IEEEcompsocnotconfonly{\sffamily}\@IEEEcompsocconfonly{\itshape}\@IEEEcompsocconfonly{\large}}
\def\@IEEEauthordefaulttextstyle{\EuMWauthorsize\@IEEEcompsocnotconfonly{\sffamily}\sublargesize}

\def\thebibliography#1{\section*{\refname}%
    \addcontentsline{toc}{section}{\refname}%
    \EuMWbibsize\@IEEEcompsocconfonly{\small}\vskip 0.3\baselineskip plus 0.1\baselineskip minus 0.1\baselineskip
    \list{\@biblabel{\@arabic\c@enumiv}}%
    {\settowidth\labelwidth{\@biblabel{#1}}%
    \leftmargin\labelwidth
    \advance\leftmargin\labelsep\relax
    \itemsep \IEEEbibitemsep\relax
    \usecounter{enumiv}%
    \let\p@enumiv\@empty
    \renewcommand\theenumiv{\@arabic\c@enumiv}}%
    \let\@IEEElatexbibitem\bibitem%
    \def\bibitem{\@IEEEbibitemprefix\@IEEElatexbibitem}%
\def\newblock{\hskip .11em plus .33em minus .07em}%
\ifCLASSOPTIONtechnote\sloppy\clubpenalty4000\widowpenalty4000\interlinepenalty100%
\else\sloppy\clubpenalty4000\widowpenalty4000\interlinepenalty500\fi%
    \sfcode`\.=1000\relax}

\long\def\@makecaption#1#2{%
\ifx\@captype\@IEEEtablestring%
\par\@IEEEtabletopskipstrut
\else
\@IEEEfigurecaptionsepspace
\fi
\setbox\@tempboxa\hbox{\normalfont\footnotesize {#1.}\nobreakspace\nobreakspace #2}%
\ifdim \wd\@tempboxa >\hsize%
\setbox\@tempboxa\hbox{\normalfont\footnotesize {#1.}\nobreakspace\nobreakspace}%
\parbox[t]{\hsize}{\normalfont\footnotesize\noindent\unhbox\@tempboxa#2}%
\else
\ifCLASSOPTIONconference \hbox to\hsize{\normalfont\footnotesize\hfil\box\@tempboxa\hfil}%
\else \hbox to\hsize{\normalfont\footnotesize\box\@tempboxa\hfil}%
\fi\fi
\ifx\@captype\@IEEEtablestring%
\@IEEEtablecaptionsepspace
\else
\fi}

\newlength\tablecaptiontotableskip
\newlength\figuretocaptionskip
\def\@IEEEfigurecaptionsepspace{\vskip\figuretocaptionskip\relax}%
\def\@IEEEtablecaptionsepspace{\vskip\tablecaptiontotableskip\relax}%

\def\abstract{\normalfont%
\@IEEEabskeysecsize\bfseries\textit{\abstractname}\,\bfseries\textit{---}\,%
\@IEEEgobbleleadPARNLSP}%

\def\IEEEkeywords{\normalfont%
\@IEEEabskeysecsize\bfseries\textit{\IEEEkeywordsname}\,\bfseries\textit{---}\,%
\@IEEEgobbleleadPARNLSP}%
\def\endIEEEkeywords{\relax\vspace{0.67ex}%
\par\if@twocolumn\else\endquotation\fi%
\normalsize\normalfont}%

\DeclareRobustCommand*{\EuMWauthorrefmark}[1]{\raisebox{0pt}[0pt][0pt]{\textsuperscript{#1}}}%
\def\@IEEEauthorblockNtopspace{0ex}
\def\@IEEEauthorblockAtopspace{1mm}
\def\IEEEkeywordsname{Keywords}
\def\subsubsection{\@startsection{subsubsection}{3}{\z@}{1.5ex plus 1.5ex minus 0.5ex}%
{0.7ex plus .5ex minus 0ex}{\normalfont\normalsize\itshape}}%
\newlength{\CPheadmatchindent}%
\def\@seccntformat#1{\hbox to\CPheadmatchindent{\csname the#1dis\endcsname}\hskip 0.1em \relax}
\IEEEilabelindentA \parindent
\IEEEilabelindent \IEEEilabelindentA
\IEEEelabelindent \parindent
\IEEEdlabelindent \parindent
\IEEElabelindent \parindent
\makeatother

\begin{document}
\raggedbottom
%
%
%
\title{Temporal Graph Neural Network for ISAC Target Detection and Tracking}
%
%
\author{%
\IEEEauthorblockN{%
Saiedeh Maboud Sanaie\EuMWauthorrefmark{\#1}, 
Marcus Großmann\EuMWauthorrefmark{\$2}, 
Markus Landmann\EuMWauthorrefmark{\$3},
Thomas Dallmann\EuMWauthorrefmark{\#4}
}
\IEEEauthorblockA{%
\EuMWauthorrefmark{\#}Electronic Measurement and Signal Processing Group, Ilmenau University of Technology, Germany\\
\EuMWauthorrefmark{\$}Fraunhofer IIS, Ilmenau, Germany\\
\{\EuMWauthorrefmark{1}saiedeh.maboud-sanaie, \EuMWauthorrefmark{4}thomas.dallmann\}@tu-ilmenau.de,
\{\EuMWauthorrefmark{2}marcus.grossmann , \EuMWauthorrefmark{3}markus.landmann\}@iis.fraunhofer.de\\
}
}
%
\maketitle
%
%
%
\begin{abstract}
Integrated sensing and communication (ISAC) is a key enabler of 6G, supporting environment-aware services. A fundamental sensing task in this setting is reliable multi-target detection and tracking. This paper proposes a temporal graph neural network (TGNN)-based tracking method that exploits delay and Doppler information from the wireless channel. The delay-Doppler map is modeled as a sequence of graphs, and tracking is formulated as a temporal node classification problem, enabling joint clustering and data association of dynamic targets. Using ray-tracing-based channel outputs as ground truth, the method is evaluated across multiple scenes with varying target positions, velocities, and trajectories and is compared with a Kalman filter baseline. Results demonstrate reduced normalized mean squared error (NMSE) in delay and Doppler, leading to more accurate multi-target tracking.
\end{abstract}
%
%
\section{Introduction}
Integrated sensing and communication (ISAC) is expected to transform 6G from a system providing pure communication functionality into a perceptive network. ISAC supports a wide range of applications, such as autonomous driving, UAV detection, and environmental monitoring~\cite{b1}. These applications often require reliable detection and tracking in multi-target scenarios. Nevertheless, sensing in multi-target scenarios remains challenging and has not yet been extensively researched in ISAC.
Conventional tracking approaches typically follow a multi-stage pipeline in which kinematic measurements are extracted from detections, clustered, associated with existing tracks via a cost-based procedure, and then processed by a tracking filter using prediction-update steps and track management \cite{b2}.
In~\cite{b3}, a practical 5G ISAC implementation for detecting and localizing multiple moving targets is proposed, but the tracking stage still relies on conventional post-detection nearest-neighbor association. In \cite{b4}, an Extended Kalman Filter (EKF)-based framework is used to investigate multi-target tracking in a monostatic ISAC system; however, the work primarily focuses on beamforming design and considers an angle-tracking scenario.
Given the complexity of conventional tracking modules, it is of interest to investigate whether learning-based methods can 
perform parts of the pipeline or even enable end-to-end tracking directly from sensing information. For instance, \cite{b5} proposes a CNN-based encoder-decoder framework for joint target detection and delay-Doppler estimation. The work in \cite{b6} employs neural-network-based appearance features for data association, whereas the tracking stage remains based on a standard Kalman filter.
CNNs combined with recurrent units is explored for radar-based tracking, particularly in single-target indoor localization \cite{b7}. Unlike CNNs, which mainly capture local patterns, Graph neural networks (GNNs) better address generalization and scalability by modeling dependencies over irregular data structures~\cite{b8}. Prior graph-based approaches already demonstrate this potential. For example, \cite{b9} formulates multi-frame detection as a graph-based link prediction problem by connecting detections from successive frames for tracking, while \cite{b10} models sequential range-Doppler maps as graphs to capture spatial and temporal correlations, using attention-based edge refinement and a learned adjacency matrix for sea-clutter suppression. Nevertheless, these methods do not explicitly account for the temporal evolution of the graph structure itself, which is central to multi-target tracking in dynamic ISAC scenarios. This motivates the use of temporal graph neural networks (TGNNs), which extend graph-based modeling to dynamic settings. 

In this work, we adopt EvolveGCN as a representative TGNN model \cite{b11} for bi-static ISAC multi-target tracking. Based on this representation, tracking is formulated as a temporal node-classification task on graphs constructed from delay-Doppler maps. The proposed method jointly learns clustering and data association over time to support simultaneous tracking of multiple targets. We evaluate the proposed method against a linear Kalman Filter tracking benchmark and assess its tracking performance through simulations.


\section{System Model}
Consider a bi-static ISAC scenario consisting of a static transmitter (Tx), a receiver (Rx), and multiple moving targets. 
The channel impulse response (CIR) is modeled as the superposition of static channel components and target-related propagation paths. Each propagation path $l$ is characterized by
a propagation delay $\tau_l$, Doppler shift $\nu_{l}$, angle of departure $(\phi_t, \theta_t)$, angle of arrival
$(\phi_r, \theta_r)$, and complex path gain $\alpha_l$. 
In this work, the CIR is generated using the Sionna ray-tracing tool.
Target-related paths are modeled following the framework in \cite{b12}
and the time-varying CIR is expressed as

\begin{equation}
    h(\tau, t) = \sum_{l=1}^{L} \alpha_l
    e^{j 2 \pi \nu_{l} t}
    \delta(\tau - \tau_l),
    \label{eq4}
\end{equation}

For a target-related path $\hat{l}$, $\alpha_{\hat{l}}$ incorporates Tx and Rx antenna radiation patterns, propagation path loss, and the radar cross section (RCS) of the target. The bi-static RCS relies on the assumption that the target is a point scatterer \cite{b13}. 
We consider an OFDM-based transmission with time-varying channel frequency response $H(f,t)=\mathcal{F}_{\tau}\{h(\tau,t)\}$. For target tracking, the channel is processed over observation windows of $N_{\mathrm{sym}}^{\mathrm{win}}$ consecutive OFDM symbols, where the $k$-th window is defined as
\begin{equation}
\mathcal{W}^k \triangleq \{\,kP,\;kP+1,\;\dots,\;kP+N_{\mathrm{sym}}^{\mathrm{win}}-1\,\},
\label{eq5}
\end{equation}
with gap size $P$ in OFDM symbols; overlapping windows occur for $P<N_{\mathrm{sym}}^{\mathrm{win}}$. The discrete-time step is given by $k=0,1,\dots,K-1$, where $K$ denotes the total number of windows over the tracking duration. The delay-Doppler map is achieved by
\begin{equation}
H_{\mathrm{DD}}(\tau,\nu)
=
\mathrm{FFT}_{t}\!\left(\mathrm{IFFT}_{f}\!\left(H(f,t)\right)\right).
\label{eq6}
\end{equation}

Based on the delay-Doppler map, a graph is constructed at each time step for the proposed tracking framework.

\section{Proposed Temporal Graph Neural Network Target Tracking}

TGNNs are designed to model graph-structured data that evolves over time. In this work, we employ EvolveGCN, a discrete-time TGNN that combines a graph convolutional network (GCN) with a gated recurrent unit (GRU) to capture temporal dependencies across consecutive time steps~\cite{b11}. At each time step \(k\), the data are represented as an undirected attributed graph \(\mathcal{G}^{k}=(\mathcal{V}^{k},\mathcal{E}^{k},\mathbf{X}^{k})\), where \(\mathcal{V}^{k}\) and \(\mathcal{E}^{k}\) denote the node and edge sets, respectively, and \(\mathbf{X}^{k}\) is the node feature matrix. The graph structure can be equivalently represented by the adjacency matrix \(\mathbf{A}^{k}\in\mathbb{R}^{N_k\times N_k}\), where \(N_k=|\mathcal{V}^{k}|\). 

Given the adjacency matrix, node features, and learnable weights ${\mathbf{W}}_{l}$, the node embeddings ${\mathbf{H}}^{k}_{l}$ are updated as

\begin{equation}
\mathbf{H}^{k}_{l+1} = \sigma\!\left((\tilde{\mathbf{D}}^{k})^{-1/2}\,\tilde{\mathbf{A}}^{k}\,(\tilde{\mathbf{D}}^{k})^{-1/2}\,\mathbf{H}^{k}_{l}\mathbf{W}_{l}\right),
\label{eq9}
\end{equation}

where \(\tilde{\mathbf{A}}^{k}=\mathbf{A}^{k}+\mathbf{I}\) denotes the adjacency matrix with self-loops, \(\tilde{\mathbf{D}}^{k}\) is the corresponding degree matrix, \(\sigma(\cdot)\) is an activation function, and \(\mathbf{W}_{l}\) is the trainable weight matrix of layer \(l\). The initial embedding matrix is given by the node features. 
The final-layer node representations $\mathbf{H}^{k}_{L}$ are fed into a multilayer perceptron (MLP) decoder, which transforms each node embedding into class logits. The network is optimized by minimizing the node classification loss over all nodes in the graph $\mathcal{G}^{k}$:
\[
\mathcal{L} = \sum_{v \in \mathcal{V}^{k}} \mathcal{J}(y_v^{k} , \hat{y}_v^{k}),
\qquad
\hat{y}_v = \mathrm{MLP}(\mathbf{H}^{k}_{L}),
\]
where $y_v$ and $\hat{y}_v$ denote the true and predicted labels of node $v$, respectively~\cite{b14}.

\begin{figure}[t]
    \centering
    \includegraphics[width=0.49\textwidth]{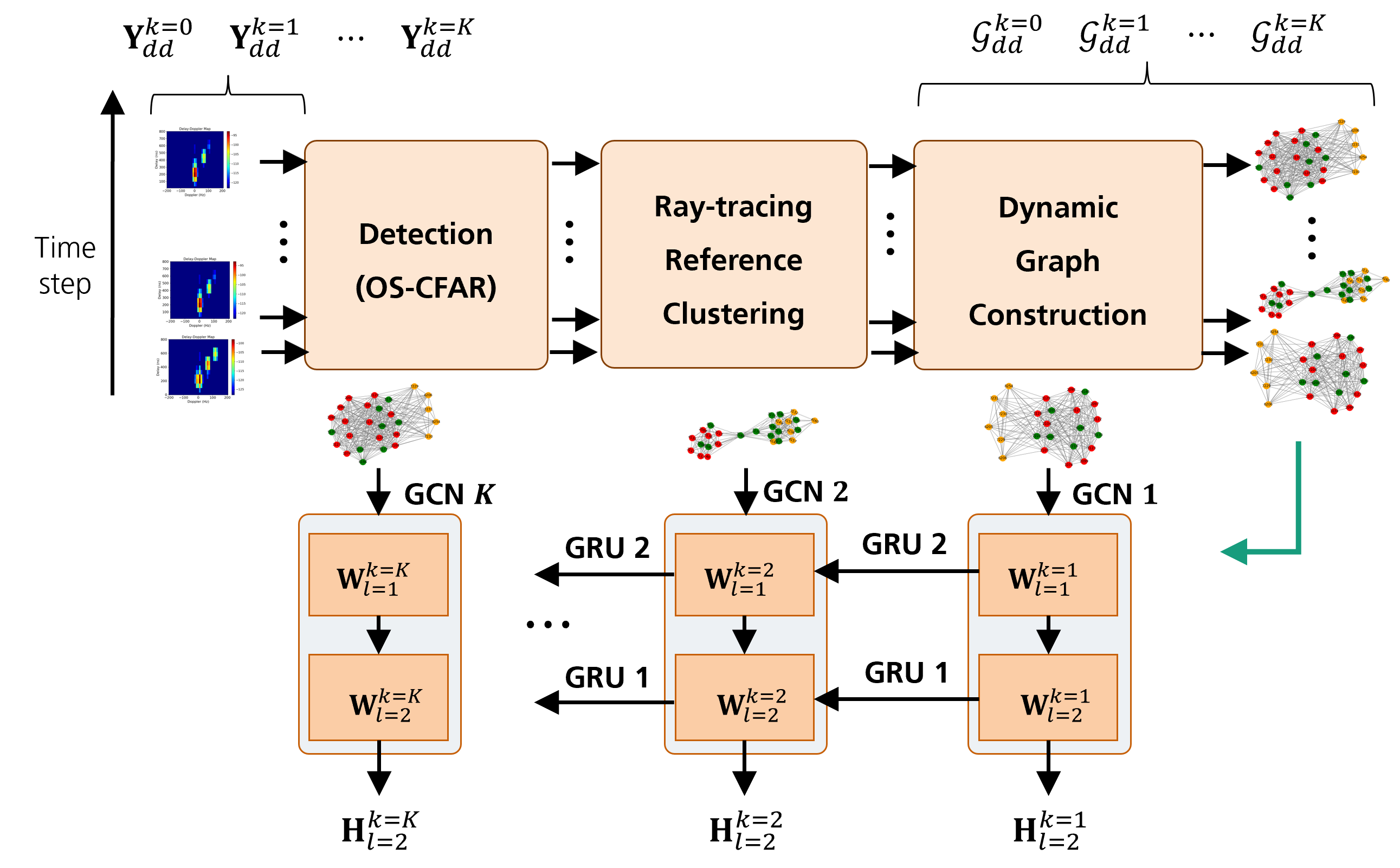}
    \caption{Framework for the proposed EvolveGCN target detection and tracking.}
    \label{fig1}
\end{figure}

Since target detection and tracking exhibit a spatiotemporal structure in the delay-Doppler domain, where target signatures evolve over time, the delay-Doppler map is represented as a temporal graph and tracking is reformulated as a temporal node-classification task to identify and track target-related nodes. 
Fig. \ref{fig1} illustrates the overall framework. A sequence of delay-Doppler maps is first converted into graph snapshots through OS-CFAR detection. The construction of the delay-Doppler graph is described in the next section. Each node in the graph is subsequently labeled according to the ground-truth ray-tracing results. The resulting graph sequence is then fed into the EvolveGCN model, where node embeddings are recursively updated to capture the temporal evolution of the scene and enable node-level target classification. EvolveGCN, is trained on labeled graph sequences and, 
at inference time, predicts target labels for nodes in unseen graphs. 

\subsection{Delay-Doppler Graph Structure}
The delay-Doppler map is not inherently graph-structured; however, CFAR detection extracts discrete delay-Doppler bins that can be represented as graph nodes. Based on these detections, we construct at each time step $k$ a graph $\mathcal{G}_{\mathrm{dd}}^{k}=(\mathcal{V}^{k},\mathcal{E}^{k},\mathbf{X}^{k})$, where each CFAR-detected bin corresponds to a node. To ensure consistent identification of the same bin across graph snapshots, each node $v \in \mathcal{V}^k$ is assigned a unique identifier given by
\begin{equation}
\mathrm{ID}_v = \ell_v^k N_{\nu} + p_v^k,
\label{eq11}
\end{equation}
where $\ell_v^k$ and $p_v^k$ denote the delay-bin and Doppler-bin indices of node $v$ at time step $k$, respectively, and $N_{\nu}$ denotes the total number of Doppler bins.
Edges $\mathcal{E}^{k}$ connect nodes that are close in delay and Doppler, so that the graph captures local relationships among detections in the delay-Doppler domain. A pair $(u,v)\in\mathcal{E}^{k}$ is created if
\begin{equation}
\big| \tau_{v}^{k} - \tau_{u}^{k} \big| \le \gamma_{\tau},\quad \text{and} \quad
\big| \nu_{v}^{k} - \nu_{u}^{k} \big| \le \gamma_{\nu},
\label{eq12}
\end{equation}
where $\gamma_{\tau}$ and $\gamma_{\nu}$ are predefined proximity thresholds in delay and Doppler, respectively.

\begin{figure}[t]
    \centering
    \subfloat[]{%
        \includegraphics[width=0.48\linewidth]{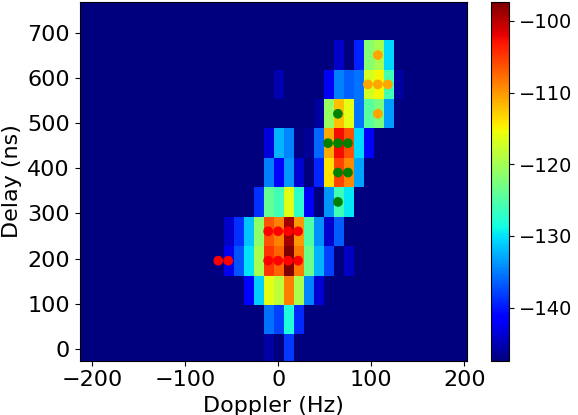}%
        \label{fig:dd_map}
    }
    \hfill
    \subfloat[]{%
        \includegraphics[width=0.48\linewidth]{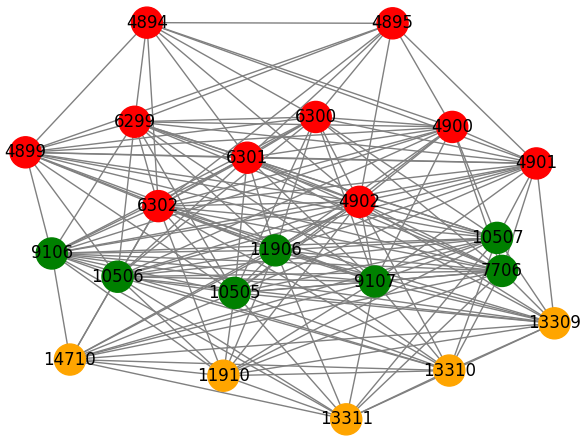}%
        \label{fig:dd_graph}
    }
    \caption{Construction of the delay-Doppler graph for window $\mathcal{W}^0$: (a) delay--Doppler map followed by CFAR and node labeling based on ray-tracing target delays and Doppler shifts, and (b) Corresponding delay-Doppler graph $\mathcal{G}_{\mathrm{dd}}^{0}$. The colors orange, red, and green correspond to target labels 0, 1, and 2, respectively.}
    \label{fig2}
\end{figure}

For supervised tracking, each node $v \in \mathcal{V}^{k}$ is assigned a target label $y_v^{k} \in \mathcal{Y}^{k}$ where $\mathcal{Y}^{k} = \{0,1,\ldots,N_{\text{target}}-1\}$, such that nodes generated by the same physical target share the same label. Fig.~\ref{fig2} illustrates the delay-Doppler map, the detected bins, and the labeled graph at time step $k=0$.
Labels correspond to target identities and are encoded as integer classes starting from zero (i.e., $0,1,2$) for three-target tracking. 

Each node $v \in \mathcal{V}^{k}$ is described by a feature vector that includes the node ID, time step, delay $\tau_v^{k}$, Doppler $\nu_v^{k}$, and power $p_v^{k}$, together with the mean delay $\bar{\tau}_{\mathcal{N}_v}^{k}$, mean Doppler $\bar{\nu}_{\mathcal{N}_v}^{k}$, and mean power $\bar{p}_{\mathcal{N}_v}^{k}$ of its neighborhood $\mathcal{N}_v^{k} \triangleq \{u \in \mathcal{V}^{k} \mid (u,v) \in \mathcal{E}^{k}\}$. The neighborhood summaries can improve the robustness and informativeness of node representations~\cite{b15}. The resulting feature vector is
\begin{equation}
\mathbf{x}_{v}^{k} =
\Big[
\mathrm{ID}_{v},\;
k,\;
\tau_{v}^{k},\;
\nu_{v}^k,\;
p_{v}^{k},\;
\bar{\tau}_{\mathcal{N}_{v}}^{k},\;
\bar{\nu}_{\mathcal{N}_{v}}^{k},\;
\bar{p}_{\mathcal{N}_{v}}^{k}
\Big]
\label{eq14}
\end{equation}



\section{Simulation Results}
We compare the proposed EvolveGCN tracker with a Kalman-filter baseline. At each time step $k$, an ordered-statistics CFAR (OS-CFAR) detector identifies target bins \cite{b16}, which are clustered using DBSCAN \cite{b17}, associated via Global Nearest Neighbor, and then processed by a linear Kalman filter for trajectory estimation \cite{b2}.
Simulations are conducted using Sionna ray tracing of a selected area of the University of Ilmenau campus. Table~\ref{tab1} summarizes the key simulation and processing parameters used in the simulation. Assuming perfect channel knowledge, we compute the delay-Doppler map and construct the corresponding graph $\mathcal{G}_{\mathrm{dd}}^{k}$ 
for each window $\mathcal{W}^k$. 
We employ EvolveGCN with a two-layer GCN ($64$ and $32$ hidden units), a history window of $6$ time steps, cross-entropy loss, and Adam with learning rate $10^{-3}$. The temporal sequence is split into $65\%$ training, $10\%$ validation, and $25\%$ test data.
\begin{table}[t]
\centering
\caption{System and target parameters.}
\label{tab1}
\begin{tabular}{>{\raggedright\arraybackslash}p{3.4cm} p{1.8cm} p{2.2cm}}
\toprule
\multicolumn{3}{l}{\textbf{System Parameters}} \\
\midrule
Parameter & Symbol & Value \\
\midrule
Subcarrier spacing & $\Delta f$ & $15\,\mathrm{kHz}$ \\
Number of subcarriers & $N_{\mathrm{FFT}}$ & $1024$ \\
OFDM symbols per window & $N_{\mathrm{sym}}^{\mathrm{win}}$ & $1400$ \\
Doppler resolution & $\Delta f_D$ & $10.75\,\mathrm{Hz}$ \\
Delay resolution & $\Delta \tau$ & $\approx 65.1\,\mathrm{ns}$ \\
Carrier frequency & $f_c$ & $5\,\mathrm{GHz}$ \\
Edge threshold & $\gamma_{\text{delay}}, \gamma_{\text{Doppler}}$ & $9$ bins \\
Tracking time & $T$ & $15\,\mathrm{s}$ \\
\end{tabular}
\vspace{6pt}
\begin{tabular}{p{1.3cm} p{1.5cm} p{1.5cm} p{2.8cm}}
\toprule
\multicolumn{4}{l}{\textbf{Target Parameters}} \\
\midrule
Target & $\mathrm{RCS}_{\min}$ & $\mathrm{RCS}_{\max}$ & Velocity vector (km/h) \\
\midrule
Target 1 & $-1.36$ & $33.98$ & $\mathbf{v}_1=[-10,\,-10,\,0]$ \\
Target 2 & $3.44$  & $32.97$ & $\mathbf{v}_2=[10,\,10,\,0]$ \\
Target 3 & $3.85$  & $7.54$  & $\mathbf{v}_3=[0,\,-10,\,0]$ \\
\bottomrule
\end{tabular}
\end{table}

\subsection{Tracking Performance Evaluation}
\begin{figure}[t]
    \centering
    \subfloat[Delay tracking performance.\label{fig3a}]{
        \includegraphics[width=0.83\linewidth]{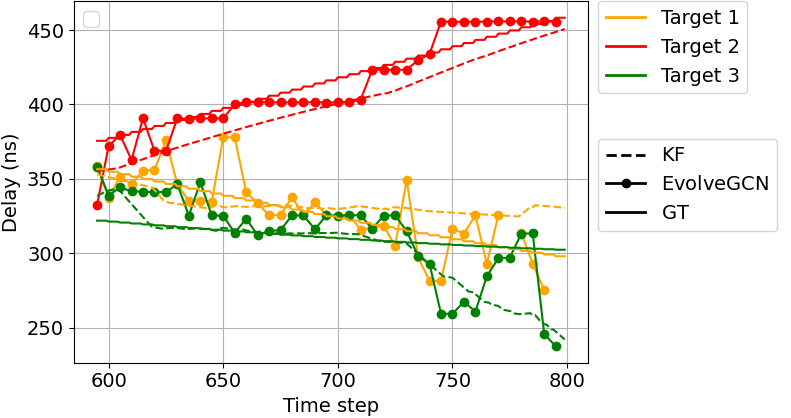}
    }

    \vspace{4pt}

    \subfloat[Doppler tracking performance.\label{fig3b}]{
        \includegraphics[width=0.85\linewidth]{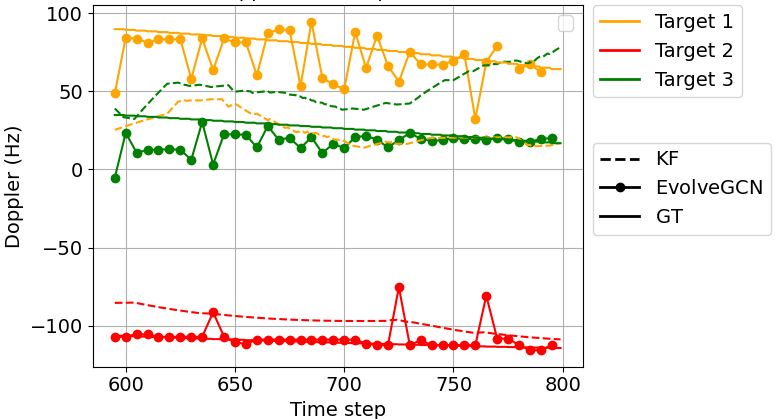}
    }
    \caption{Tracking performance over the test time steps: EvolveGCN predictions versus ground truth (GT) and Kalman filter (KF).}
    \label{fig3}
\end{figure}

In Fig.~\ref{fig3a} and~\ref{fig3b}, delay and Doppler tracking results of Kalman filter and EvolveGCN are compared against ground-truth values obtained 
from ray tracing over the test time steps for each target. 
For the Kalman filter baseline, the state is initialized at $k=0$ with the ground-truth target state. 
A key difference is that the Kalman filter can still output state estimates when observations are missing through its prediction step, whereas a graph-based method 
typically cannot provide a meaningful estimate if the target corresponding node is absent. 
For Target~1, the corresponding node is intermittently missing during the test interval; therefore, the tracking error is undefined at those time steps. 
We evaluate the Root Mean Square Error (RMSE) of delay and Doppler, with the error defined as
\begin{equation}
e_{\tau}(k)=\bigl|\hat{\tau}(k)-\tau_{\mathrm{gt}}(k)\bigr|,\qquad
e_{\nu}(k)=\bigl|\hat{\nu}(k)-\nu_{\mathrm{gt}}(k)\bigr|,
\end{equation}


\begin{figure}[!t]
    \centering
    \includegraphics[width=1\linewidth]{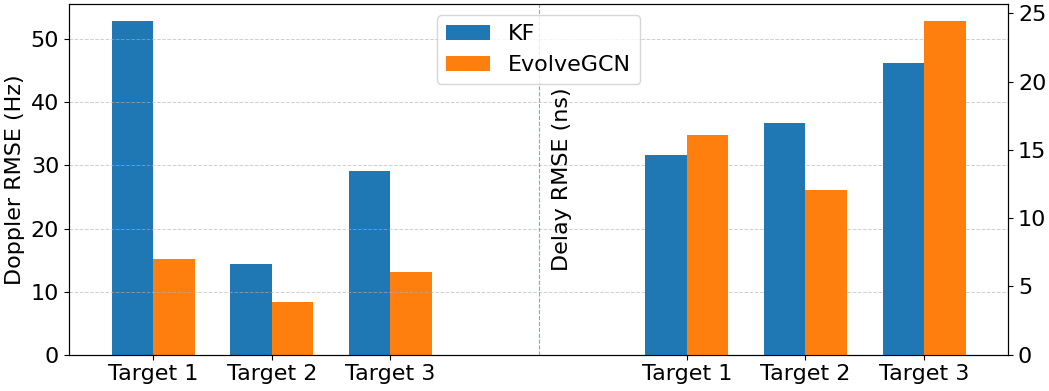}
    \caption{RMSE of Doppler and delay tracking per target for the Kalman filter and EvolveGCN.}
    \label{fig5}
\end{figure}

The RMSE of the delay and Doppler tracking errors for both schemes is shown in Fig.~\ref{fig5}. For Doppler tracking, the Kalman filter achieves an RMSE of less than 5 bins at a resolution of 10.75~Hz per bin, whereas EvolveGCN reduces it to less than 2 bins. For delay tracking, both methods achieve sub-bin accuracy at a resolution of 65.1~ns per bin. 

Overall, EvolveGCN provides notably better Doppler tracking performance.
We evaluated robustness and generalization by running EvolveGCN and a Kalman filter over four independent scenes, while keeping the tracking duration and all other simulation parameters fixed. 
Across scenes, we randomized the initial positions of the Tx, Rx, and targets, as well as the target velocity vectors. Consequently, each scene defines a newly generated target trajectory under 
a constant-velocity motion model, with target speeds in the range of 
10-15~km/h. For each scene $s \in {S}$, target $c \in {C}$, and test time step $k \in {K}$, both methods produced delay and Doppler tracking estimates 
$\hat{x}_{s,c,k}\in\{\hat{\tau}_{s,c,k},\, \hat{{\nu}}_{s,c,k}\}$, which were compared against the ray-tracing ground truth $x_{s,c,k}\in\{\tau_{s,c,k},\, {\nu}_{s,c,k}\}$. 
The performance is reported in terms of Normalised Mean Square Error 
(NMSE), computed by averaging the squared estimation errors over all scenes, targets, and time steps: 
\begin{equation}
\mathrm{NMSE}=\frac{\sum_{s=1}^{S}\sum_{c=1}^{C}\sum_{k=0}^{K-1}
m_{s,c,k}\,\bigl(\hat{x}_{s,c,k}-x_{s,c,k}\bigr)^{2}}{\sum_{s=1}^{S}\ \sum_{c=1}^{C}\ \sum_{k=0}^{K}
m_{s,c,k}\left(x_{s,c,k}\right)^2}
\end{equation}
Here, $m_{s,c,k}$ denotes a binary mask. It is set to $1$ if both the prediction $\hat{x}_{s,c,k}$ and the corresponding ground-truth value $x_{s,c,k}$ are available, and to $0$ otherwise. Thus, the error is computed only over valid prediction--ground-truth pairs.
As shown in Table~\ref{tab3}, EvolveGCN achieves a $40.2\%$ NMSE reduction for delay and a $68.4\%$ reduction for Doppler compared to Kalman filtering.
\begin{table}[!t]
\centering
\caption{NMSE results for delay and Doppler tracking.}
\label{tab:nmse_results}
\begin{tabular}{lcc}
\hline
\textbf{Method} & $\mathbf{\mathrm{NMSE}_{\tau}}$ & $\mathbf{\mathrm{NMSE}_{\nu}}$ \\
\hline
Kalman Filter & $0.0665$ & $0.288$ \\
EvolveGCN & $0.0398$ & $0.0910$ \\
\hline
\end{tabular}
\label{tab3}
\end{table}




\section{Conclusion}

In this paper, we formulate bi-static ISAC multi-target tracking as a temporal node-classification problem on a sequence of delay-Doppler graphs and use EvolveGCN to infer time-varying target labels. Compared with a Kalman-filter baseline, the proposed approach achieves lower NMSE in delay and Doppler over multiple scenes; future work will incorporate closed-loop feedback for adaptive beamforming, as well as practical channel estimation and noisy measurements.

\section*{Acknowledgement}
This work was developed within the GKS-6G project,
Supported by the Free State of Thuringia and the European Social Fund Plus under grant 2024 FGR 0061.


\bibliographystyle{IEEEtran}
\bibliography{IEEEabrv, references}

\end{document}